\documentclass[reprint, 
groupedaddress,
superscriptaddress,
longbibliography,
amsmath,amssymb,
aps,
prx,
]{revtex4-2}

\usepackage[colorlinks=true,linkcolor=blue,citecolor=blue,urlcolor=blue]{hyperref}
\usepackage{graphicx}
\usepackage{dcolumn}
\usepackage{bm}
\usepackage{xcolor}
\usepackage{braket}
\usepackage{microtype}

\newcommand{\D}{\mathcal{D}}
\newcommand{\Tr}{\operatorname{Tr}}

\begin{document}

\title{Collectively Enhanced Universal Photon Blockade}

\author{Guohao Chang}
\affiliation{School of Physics and Electronic Engineering, Xinjiang Normal University, Urumqi, Xinjiang 830054, China}

\author{Ahmad Abliz}
\email{aahmad@126.com}
\affiliation{School of Physics and Electronic Engineering, Xinjiang Normal University, Urumqi, Xinjiang 830054, China}

\date{\today}

\begin{abstract}
High-purity and bright single-photon sources are important for quantum information processing and precision measurement. We propose a collectively enhanced universal photon-blockade scheme in a multi-emitter two-photon Tavis--Cummings system. Independent coherent drives of the cavity and collective emitter enable destructive interference between two-photon excitation pathways, while collective coupling supplies level anharmonicity. Full-quantum master-equation simulations, the Holstein--Primakoff approximation, and a non-Hermitian probability-amplitude analysis yield the optimal conditions: cavity resonance together with phase and amplitude matching. Compared with a purely collective blockade, the proposed scheme lowers \(g^{(2)}(0)\), and essentially preserves the single-photon population. As the emitter number \(N\) increases, the optimal point remains at \(\Delta_c=0\), while the minimum correlation follows \(g_{\min}^{(2)}(0)\propto N^{-2}\). By contrast, unconventional photon blockade shifts away from resonance as \(N\) increases, limiting its purity improvement and brightness. In the weak-drive regime, universal blockade obeys \(g^{(2)}(0)\propto\varepsilon_a^2\), owing to a higher-order bypass through the three-excitation manifold. The drive strength therefore provides an additional purity--brightness control, although the interference mechanism makes the scheme more sensitive to dissipation-rate mismatch. Our results establish a scalable route to bright, high-purity, and tunable single-photon emission in multi-emitter cavity-QED systems.
\end{abstract}

\maketitle

\section{Introduction}

High-quality single-photon sources are fundamental resources for quantum communication, quantum computing, quantum metrology, and quantum networks \cite{1,2,3,4,5,6,7}. An ideal single-photon source should exhibit pronounced photon antibunching while simultaneously providing high emission brightness, stability, and scalability. Photon blockade (PB) is an important mechanism for producing nonclassical single-photon streams\cite{9,10,11}. Its physical picture is analogous to Coulomb blockade in electronic transport: the presence of one photon modifies the transition condition for subsequent excitations and thereby suppresses the entry of a second photon \cite{grabert2013single}. This effect is commonly characterized by the equal-time second-order correlation function \(g^{(2)}(0)\). The condition \(g^{(2)}(0)<1\) indicates photon antibunching, whereas \(g^{(2)}(0)\ll1\) corresponds to high-purity single-photon emission. Depending on the physical origin of multiphoton suppression, PB is generally classified as conventional photon blockade (CPB) or unconventional photon blockade (UPB). CPB relies on the anharmonic energy spectrum generated by light--matter interactions: the single-photon transition is resonant while the two-photon transition is off resonant, suppressing the excitation of a second photon. Resolving the anharmonic energy splitting from the system linewidth, however, generally requires the individual emitter--cavity coupling to exceed both the cavity loss rate and the emitter decay rate, imposing a demanding strong-coupling requirement\cite{11,12,13,15,16,17,18}. UPB instead suppresses the two-photon population through destructive interference between different excitation pathways and can produce strong antibunching even in the weak-coupling regime \cite{19,21,27,31,45}. Such schemes, however, usually operate away from the single-photon resonance and therefore tend to reduce the single-photon population. Moreover, destructive interference requires precise matching of pathway amplitudes and phases, making the blockade sensitive to detunings, drive parameters, and dissipation rates.

Introducing multiple emitters and exploiting collective coupling offers an important route for relaxing the individual strong-coupling requirement \cite{2019,Collective}. For a symmetric collective mode formed by \(N\) identical emitters, the effective light--matter coupling is enhanced to \(g_N=g\sqrt{N}\). In the standard one-photon Tavis--Cummings model, however, increasing the number of emitters enhances the collective Rabi splitting while making the low-excitation spectrum increasingly harmonic, thereby weakening the level anharmonicity required for PB\cite{down1,down2,down3,Collective}. The two-photon Tavis--Cummings model provides a different route around this limitation. Because its interaction exchanges two cavity photons with one emitter excitation, the nonlinearity originates from the two-photon light--matter coupling itself and remains finite even in the large-\(N\) limit. Recent studies have shown that such two-photon interactions can generate collectively enhanced CPB, for which the second-order correlation function decreases substantially with emitter number while a high single-photon population is retained \cite{Collective}. This scheme, however, mainly exploits level anharmonicity and does not use the additional suppression and tunability offered by pathway interference.

The recently proposed universal photon blockade (universal PB) provides a route for combining energy-level anharmonicity with quantum interference \cite{37,38,39,372,chang2026,dong2026,universal2026}. In a single-emitter two-photon Jaynes--Cummings system, independent driving of the cavity and emitter creates two pathways to the two-photon state: one in which two photons are generated sequentially by the cavity drive, and another in which the emitter is first excited and subsequently converted into two cavity photons through the two-photon exchange interaction. When the system simultaneously satisfies the single-photon resonance condition and the phase- and amplitude-matching conditions of the two pathways, level anharmonicity and destructive interference cooperate to suppress two-photon excitation. Compared with UPB, which completely avoids the single-photon resonance, universal PB can yield stronger antibunching while maintaining a relatively high single-photon population and can operate over a broader range of coupling strengths. Previous studies have mainly focused on universal PB with a single emitter, and the scaling behavior resulting from the combination of pathway interference and many-body collective effects remains unclear \cite{37,38,39,372,chang2026,dong2026,universal2026}. Collective coupling enhances the level anharmonicity but also changes the relative amplitudes and optimal conditions of the excitation pathways; it therefore need not benefit all interference-based blockade mechanisms equally. In particular, it is important to determine whether universal PB and UPB, which perform similarly in the single-emitter case, remain comparable in a multi-emitter system; whether universal PB can enhance purity while preserving a resonant operating point and high brightness; and how it responds to variations in drive strength and mismatched dissipation rates. These questions are central to evaluating the scalability and experimental feasibility of the mechanism.

In this work, we study a two-photon Tavis--Cummings system composed of a single-mode optical cavity and \(N\) identical two-level emitters, with independently controllable coherent drives applied to the cavity mode and the collective emitter mode. We combine full-quantum master-equation simulations exploiting permutation symmetry, the Holstein--Primakoff approximation, and a probability-amplitude method based on a non-Hermitian effective Hamiltonian, thereby cross-validating the finite-size, low-excitation, and analytical descriptions. In the weak-drive regime, we derive the optimal conditions for universal PB and show that the energy selectivity provided by cavity resonance and destructive interference between two excitation pathways can be realized simultaneously in a multi-emitter system. Our main results are as follows. First, compared with a purely collective scheme based only on collective anharmonicity, interference further reduces \(g^{(2)}(0)\) and broadens the effective blockade window without substantially changing the single-photon population. Second, universal PB exhibits clear collective enhancement: its optimal operating point remains at the cavity resonance, and its minimum second-order correlation function scales as \(g_{\min}^{(2)}(0)\propto N^{-2}\). By contrast, the optimal detuning of UPB moves away from resonance with increasing \(N\), and neither its blockade performance nor its brightness benefits equally from collective enhancement, demonstrating that collective effects act selectively on different interference mechanisms. Third, we find \(g^{(2)}(0)\propto\varepsilon_a^2\) in the weak-drive regime. By extending the probability-amplitude truncation to the three-excitation subspace, we attribute this behavior to higher-order bypass pathways that avoid the lowest-order two-photon state and lead to the three-photon state. This scaling provides an additional degree of freedom for tuning the purity--brightness relation. Finally, we examine the effect of mismatched dissipation rates and find that, although universal PB produces stronger antibunching, its reliance on precise pathway balance makes it less robust than the purely collective scheme. These results reveal the cooperation between many-body collective enhancement and two-path quantum interference and provide a route toward high-purity, bright, and tunable single-photon emission under relatively weak individual emitter coupling.

\section{System Model}
\label{sec:model}

We consider a system consisting of a single-mode optical cavity and \(N\) identical two-level emitters (atoms). Light and matter interact through a two-photon exchange process: the creation or annihilation of two cavity photons is accompanied by the de-excitation or excitation of one emitter. Such two-photon coupling has been realized in platforms including superconducting artificial atoms and quantum dots \cite{realized1,realized2,realized3,realized4,realized5,realized6,realized7,realized8,realized9,realized10}. To enable flexible control, the cavity and the emitter ensemble are driven by independent classical coherent fields. The frequency, amplitude, and phase of the cavity drive are denoted by \(\omega_p^a\), \(\varepsilon_a\), and \(\phi_a\), respectively, whereas those of the collective emitter drive are denoted by \(\omega_p^e\), \(\varepsilon_e\), and \(\phi_e\). Experimentally, the relative phase between the two drives can be independently controlled using established microwave or optical phase-control techniques \cite{phase}.

In the rotating frame defined by the unitary transformation
\begin{equation}
U(t)=\exp\left[-it\left(\omega_p^a a^\dagger a+\omega_p^e\sum_{i=1}^{N}\sigma_i^+\sigma_i^-\right)\right],
\end{equation}
the coherent dynamics are governed by the generalized two-photon Tavis--Cummings Hamiltonian
\begin{align}
H_{\mathrm{TC}}={}&\Delta_c a^\dagger a
+\Delta_e\sum_{i=1}^{N}\sigma_i^+\sigma_i^-
+g\sum_{i=1}^{N}\left(a^{\dagger 2}\sigma_i^-+a^2\sigma_i^+\right) \notag\\
&+\varepsilon_a\left(a^\dagger e^{i\phi_a}+ae^{-i\phi_a}\right) \notag\\
&+\frac{\varepsilon_e}{\sqrt N}\sum_{i=1}^{N}
\left(\sigma_i^+e^{i\phi_e}+\sigma_i^-e^{-i\phi_e}\right).
\label{eq:HTC}
\end{align}
Here, \(a\) (\(a^\dagger\)) is the annihilation (creation) operator of the cavity mode, \(\sigma_i^+=|e_i\rangle\langle g_i|\) and \(\sigma_i^-=|g_i\rangle\langle e_i|\) describe the excitation and de-excitation of the \(i\)th emitter, and \(g\) is the two-photon coupling strength between a single emitter and the cavity mode.

Throughout this work, we assume the resonance relation \(\omega_e=2\omega_c\) between the emitter and cavity frequencies, together with the corresponding drive-frequency relation \(\omega_p^e=2\omega_p^a\). Consequently, the cavity detuning \(\Delta_c=\omega_c-\omega_p^a\) and emitter detuning \(\Delta_e=\omega_e-\omega_p^e\) satisfy
\begin{equation}
\Delta_e=2\Delta_c.
\end{equation}

To describe the dissipative dynamics accurately, we solve the full-quantum master equation numerically. The primary dissipative processes are cavity-photon leakage and spontaneous emission from each emitter, with decay rates \(\gamma_c\) and \(\gamma_e\), respectively. The density operator \(\rho(t)\) obeys the Lindblad master equation
\begin{equation}
\dot\rho=-i[H_{\mathrm{TC}},\rho]
+\gamma_c\D[a]\rho
+\gamma_e\sum_{i=1}^{N}\D[\sigma_i^-]\rho,
\label{eq:master_full}
\end{equation}
where
\begin{equation}
\D[O]\rho=O\rho O^\dagger-\frac{1}{2}
\left(O^\dagger O\rho+\rho O^\dagger O\right)
\end{equation}
is the Lindblad dissipator. We focus on the steady state, obtained by solving \(\dot\rho=0\), with \(\rho_s=\lim_{t\to\infty}\rho(t)\). All numerical calculations are performed using the quantum-optics toolbox QuTiP \cite{46,47,qutip}. Because the identical emitters possess permutation symmetry, the original Hilbert-space dimension \(\mathcal O(d_c2^N)\), where \(d_c\) is the cavity photon-number cutoff, grows exponentially with \(N\). To reduce the computational complexity, we exploit permutation symmetry at the density-matrix level, reducing the number of independent matrix elements from the exponential scaling \(\mathcal O(4^N)\) to the polynomial scaling \(\mathcal O(N^3)\) \cite{Kirton2017PRL,Kirton2018Superradiant,qutip}. This allows us to study larger emitter numbers without compromising numerical accuracy.

Once the steady-state density operator is obtained, the equal-time second-order correlation function is calculated as
\begin{equation}
g^{(2)}(0)=
\frac{\langle a^\dagger a^\dagger aa\rangle}
{\langle a^\dagger a\rangle^2}
=\frac{\Tr\!\left(a^\dagger a^\dagger aa\rho_s\right)}
{\left[\Tr\!\left(a^\dagger a\rho_s\right)\right]^2}.
\label{eq:g2}
\end{equation}
The condition \(g^{(2)}(0)<1\) indicates sub-Poissonian photon statistics and photon antibunching. A smaller \(g^{(2)}(0)\) corresponds to a higher single-photon purity. Experimentally, the equal-time second-order correlation function can be measured using a Hanbury Brown--Twiss setup \cite{HBT1956,48}.

\subsection{Holstein--Primakoff Approximation}
\label{sec:hp}

To obtain analytical insight and reveal the collective-enhancement mechanism in the large-\(N\) limit, we bosonize the emitter ensemble using the Holstein--Primakoff transformation. In the weak-drive regime relevant to PB, the number of collective excitations satisfies \(\langle b^\dagger b\rangle\ll N\), so that the low-excitation approximation is naturally fulfilled. We define the collective spin operators
\begin{equation}
J_+=\sum_{i=1}^{N}\sigma_i^+,\qquad
J_-=\sum_{i=1}^{N}\sigma_i^-,\qquad
J_z=\frac{1}{2}\sum_{i=1}^{N}\sigma_i^z.
\end{equation}
Using the spin-down convention, the Holstein--Primakoff mapping in the low-excitation regime reads
\begin{align}
J_-&\simeq\sqrt N\,b-\frac{1}{2\sqrt N}b^\dagger bb,\\
J_+&\simeq\sqrt N\,b^\dagger-\frac{1}{2\sqrt N}b^\dagger b^\dagger b,\\
J_z&=b^\dagger b-\frac{N}{2},
\end{align}
where \(b\) (\(b^\dagger\)) is the annihilation (creation) operator of the bosonic mode describing collective emitter excitations. The retained \(1/\sqrt N\) nonlinear correction accounts for finite-\(N\) saturation and vanishes in the thermodynamic limit \(N\to\infty\).

Substituting the transformation into Eq.~\eqref{eq:HTC} and neglecting a constant energy shift yields the effective Holstein--Primakoff Hamiltonian
\begin{align}
H_{\mathrm{HP}}={}&\Delta_c a^\dagger a+\Delta_e b^\dagger b
+g_N\left(a^{\dagger2}b+a^2b^\dagger\right) \notag\\
&+\varepsilon_a\left(a^\dagger e^{i\phi_a}+ae^{-i\phi_a}\right)
+\varepsilon_e\left(b^\dagger e^{i\phi_e}+be^{-i\phi_e}\right) \notag\\
&-\chi\left[(b^\dagger b)a^{\dagger2}b
+a^2b^\dagger(b^\dagger b)\right],
\label{eq:HHP}
\end{align}
where
\begin{equation}
g_N=g\sqrt N,\qquad
\chi=\frac{g}{2\sqrt N}.
\end{equation}
Here, \(g_N\) is the collectively enhanced coupling strength, and \(\chi\) is the nonlinear saturation coefficient associated with a finite number of emitters. Importantly, unlike the case of linear coupling, the nonlinearity in the two-photon model originates from the light--matter interaction itself rather than from emitter saturation. Therefore, even though the saturation correction vanishes as \(N\to\infty\), the spectrum remains intrinsically anharmonic, allowing PB to survive and become collectively enhanced in the thermodynamic limit.

Within the Holstein--Primakoff approximation, the master equation becomes
\begin{equation}
\dot\rho=-i[H_{\mathrm{HP}},\rho]
+\gamma_c\D[a]\rho+\gamma_e\D[b]\rho.
\label{eq:master_hp}
\end{equation}
The simplified model agrees with the full-quantum model in the low-excitation regime while substantially reducing the computational cost. It forms the basis for deriving the optimal blockade conditions and studying the large-\(N\) scaling behavior.

\section{Results and Discussion}
\label{sec:results}

\subsection{Energy Spectrum}

\begin{figure}[t]
  \centering
  \includegraphics[width=1\linewidth]{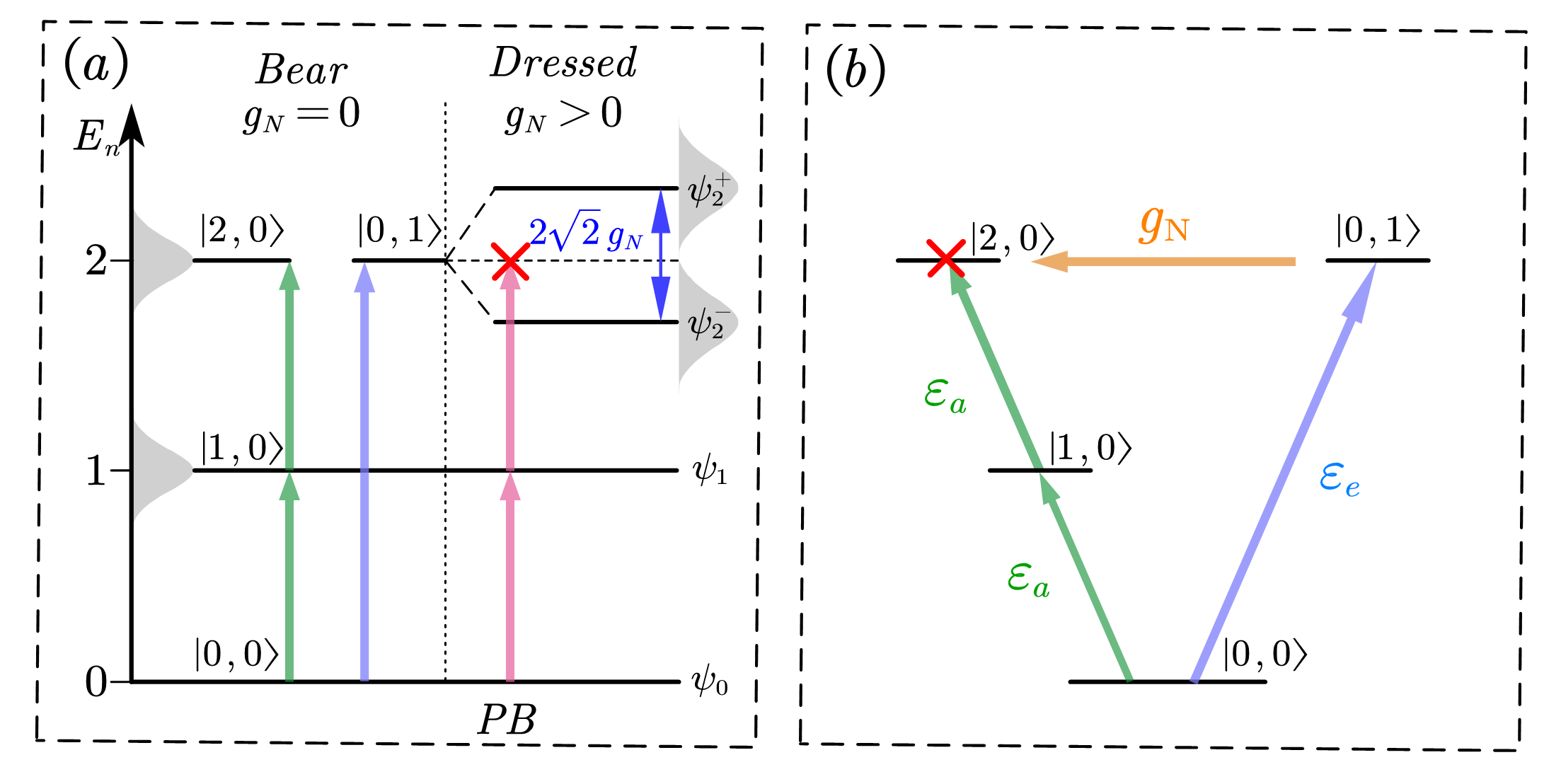}
  \caption{\textbf{Energy-level structure and interference mechanism.}
  \textbf{(a)} Conventional photon blockade (CPB) produced by the single-photon resonance and the anharmonic spectrum of the two-photon Tavis--Cummings model.
  \textbf{(b)} Two excitation pathways leading to the two-photon state \(|2,0\rangle\) and their destructive interference.}
  \label{fig:mechanism}
\end{figure}

Neglecting the drive terms (\(\varepsilon_a=\varepsilon_e=0\)) and the saturation correction (\(\chi=0\)), the Hamiltonian reduces to
\begin{equation}
H=\Delta_c a^\dagger a+\Delta_e b^\dagger b
+g_N\left(a^{\dagger2}b+a^2b^\dagger\right).
\label{eq:H_undriven}
\end{equation}
The total excitation number
\begin{equation}
N_{\mathrm{ex}}=a^\dagger a+2b^\dagger b
\end{equation}
is conserved. In the laboratory frame and under the light--matter resonance condition \(\omega_e=2\omega_c\), the eigenvalues and eigenstates in the zero-, one-, and two-excitation subspaces are
\begin{align}
E_0&=0,
&|\psi_0\rangle&=|0,0\rangle,\\
E_1&=\omega_c,
&|\psi_1\rangle&=|1,0\rangle,\\
E_2^\pm&=2\omega_c\pm\sqrt2g_N,
&|\psi_2^\pm\rangle&=\frac{1}{\sqrt2}
\left(|2,0\rangle\pm|0,1\rangle\right).
\end{align}
As illustrated in Fig.~\ref{fig:mechanism}(a), under the single-photon resonance condition the system is resonantly excited from the ground state \(|\psi_0\rangle\) to the single-photon state \(|\psi_1\rangle\). The two-photon interaction, however, shifts the two dressed levels by \(\pm\sqrt2g_N\), giving a total splitting \(2\sqrt2g_N\). The transitions from \(|\psi_1\rangle\) to the two-excitation states \(|\psi_2^\pm\rangle\) are consequently off resonant. The entrance of a second photon is therefore suppressed, giving rise to conventional photon blockade (CPB). Because the collective coupling scales as \(g_N=g\sqrt N\), increasing the number of emitters can generate a sufficiently strong effective nonlinearity even when the individual coupling \(g\) is weak, thereby relaxing the individual emitter--cavity coupling requirement of CPB.

\subsection{Collectively Enhanced Universal Photon Blockade}
\label{sec:universal}

To derive analytical results for universal PB and validate the numerical simulations, we employ a probability-amplitude approach. In the weak-drive limit, the excitation number remains small, and the wavefunction can be truncated to the subspace with \(N_{\mathrm{ex}}\leq2\). According to the quantum-trajectory description, the non-Hermitian effective Hamiltonian is
\begin{equation}
H_{\mathrm{eff}}=H_{\mathrm{HP}}
-\frac{i\gamma_c}{2}a^\dagger a
-\frac{i\gamma_e}{2}b^\dagger b.
\end{equation}
The saturation correction does not contribute within this truncation, and the wavefunction can be approximated as
\begin{equation}
|\psi\rangle=C_{00}|0,0\rangle+C_{10}|1,0\rangle
+C_{01}|0,1\rangle+C_{20}|2,0\rangle.
\end{equation}
Substituting this ansatz into the non-Hermitian Schrödinger equation \(i\partial_t|\psi\rangle=H_{\mathrm{eff}}|\psi\rangle\), imposing the steady-state condition \(\dot C_{ij}=0\), and taking \(C_{00}\simeq1\), we obtain
\begin{align}
\widetilde\Delta_c C_{10}
+\varepsilon_a e^{i\phi_a}
+\sqrt2\varepsilon_a e^{-i\phi_a}C_{20}&=0,\\
\widetilde\Delta_e C_{01}
+\varepsilon_e e^{i\phi_e}
+\sqrt2g_NC_{20}&=0,\\
\widetilde\Delta_{2c}C_{20}
+\sqrt2g_NC_{01}
+\sqrt2\varepsilon_a e^{i\phi_a}C_{10}&=0,
\end{align}
where
\begin{equation}
\widetilde\Delta_c=\Delta_c-\frac{i\gamma_c}{2},\qquad
\widetilde\Delta_e=\Delta_e-\frac{i\gamma_e}{2},\qquad
\widetilde\Delta_{2c}=2\Delta_c-i\gamma_c.
\end{equation}
Retaining the lowest nonvanishing orders in the weak drives gives
\begin{align}
C_{10}&=-\frac{\varepsilon_a e^{i\phi_a}}{\widetilde\Delta_c},\\
C_{01}&=-\frac{\varepsilon_e e^{i\phi_e}}{\widetilde\Delta_e},\\
C_{20}&=\sqrt2\,
\frac{g_N\varepsilon_e e^{i\phi_e}\widetilde\Delta_c
+\varepsilon_a^2e^{2i\phi_a}\widetilde\Delta_e}
{\widetilde\Delta_c\widetilde\Delta_e\widetilde\Delta_{2c}}.
\label{eq:C20}
\end{align}
The steady-state second-order correlation function is approximately
\begin{equation}
g^{(2)}(0)\simeq\frac{2|C_{20}|^2}{|C_{10}|^4}.
\end{equation}

Assuming \(\gamma_c=\gamma_e\) and imposing the single-photon resonance condition \(\Delta_c=\Delta_e=0\), the requirement \(C_{20}=0\) yields the optimal conditions for universal PB:
\begin{equation}
\Delta_c=0,\qquad
\phi_e=2(\phi_a+k\pi),\quad k\in\mathbb Z,\qquad
\varepsilon_e=-\frac{\varepsilon_a^2}{g_N}.
\label{eq:optimal_universal}
\end{equation}
Here, the drive amplitude \(\varepsilon_e\) is treated as a signed quantity. Equivalently, the minus sign may be absorbed into the relative phase between the two drives.

The physical mechanism of destructive interference is illustrated in Fig.~\ref{fig:mechanism}(b). There are two dominant pathways from the ground state to the two-photon state: the direct pathway
\begin{equation}
|0,0\rangle\xrightarrow{\varepsilon_a}|1,0\rangle
\xrightarrow{\varepsilon_a}|2,0\rangle
\end{equation}
and the indirect pathway
\begin{equation}
|0,0\rangle\xrightarrow{\varepsilon_e}|0,1\rangle
\xrightarrow{g_N}|2,0\rangle.
\end{equation}
When the two pathways satisfy Eq.~\eqref{eq:optimal_universal}, their probability amplitudes interfere destructively and the two-photon state is strongly suppressed. The single-photon resonance favors occupation of the single-photon state, while the matched amplitudes and phases cancel the two-photon excitation pathways. Their cooperative action produces stronger antibunching than either mechanism alone.

\begin{figure}[t]
  \centering
  \includegraphics[width=1\linewidth]{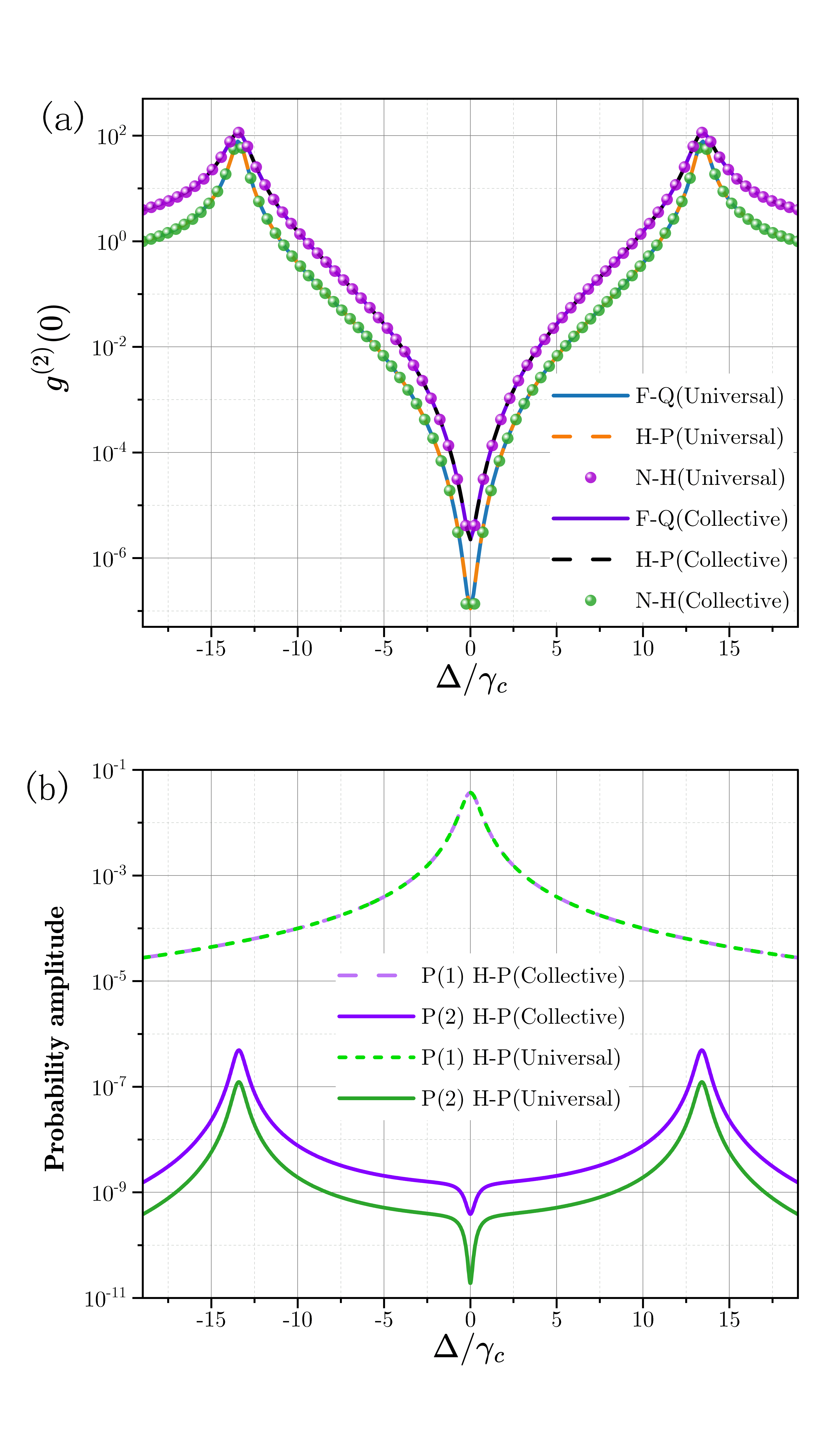}
  \caption{\textbf{Comparison between collectively enhanced universal photon blockade (universal PB) and purely collective photon blockade (purely collective PB).}
  \textbf{(a)} Equal-time second-order correlation function \(g^{(2)}(0)\) as a function of the detuning \(\Delta_c/\gamma_c\). Results from the Holstein--Primakoff approximation, full-quantum simulations, and the non-Hermitian probability-amplitude method agree closely.
  \textbf{(b)} Single- and two-photon populations as functions of detuning. The parameters are \(g/\gamma_c=6\), \(\varepsilon_a/\gamma_c=0.1\), \(\varepsilon_e=-\varepsilon_a^2/g_N\), \(\gamma_c=\gamma_e\), and \(N=10\).}
  \label{fig:collective_comparison}
\end{figure}

Figure~\ref{fig:collective_comparison}(a) compares the collectively enhanced universal PB with the purely collective PB that does not rely on pathway interference. Each scheme is evaluated independently using full-quantum simulations, numerical simulations of the Holstein--Primakoff model, and the non-Hermitian probability-amplitude solution. The three methods agree closely over the investigated parameter range, confirming the validity of the Holstein--Primakoff approximation and the analytical treatment. Compared with the purely collective scheme, the universal scheme not only substantially lowers the minimum of \(g^{(2)}(0)\), but also broadens the detuning interval over which strong antibunching occurs.

As shown in Fig.~\ref{fig:collective_comparison}(b), introducing interference to improve the blockade leaves the single-photon population of the universal scheme nearly identical to that of the cavity-only driven collective scheme. This result can also be understood analytically. Under the optimal conditions,
\begin{equation}
P_1\simeq|C_{10}|^2\propto\frac{\varepsilon_a^2}{\gamma_c^2}
\end{equation}
is independent of the collective coupling strength \(g_N\). The interference mechanism therefore introduces no additional loss of single-photon brightness. The proposed scheme can thus further enhance the single-photon purity while maintaining a high single-photon population.

\begin{figure}[t]
  \centering
  \includegraphics[width=1\linewidth]{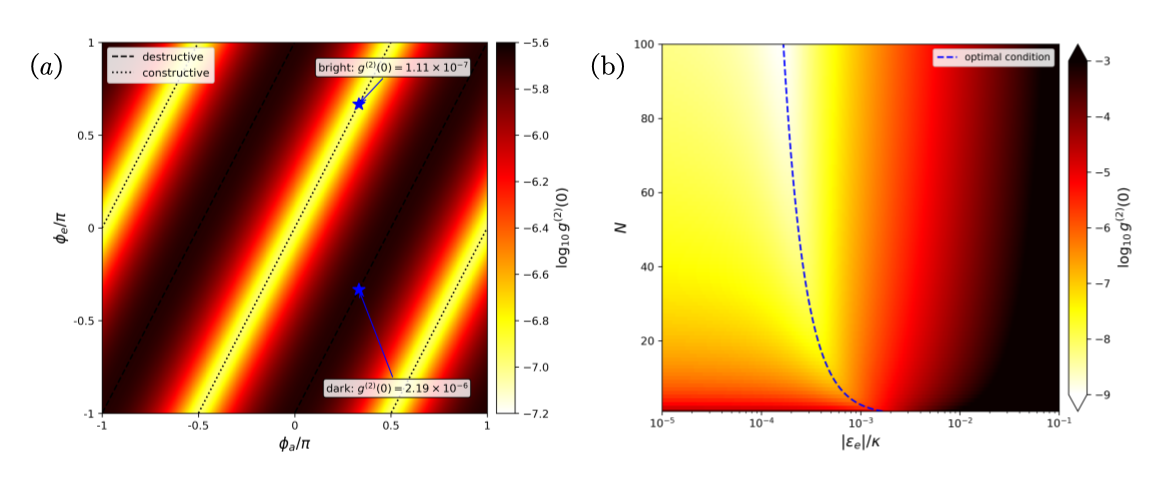}
  \caption{\textbf{Control of universal photon blockade (universal PB) through the drive phases and amplitudes.}
  \textbf{(a)} Two-dimensional distribution of \(g^{(2)}(0)\) as a function of the drive phases \(\phi_a\) and \(\phi_e\). The dashed and dotted lines indicate the two phase-matching conditions shown in the panel.
  \textbf{(b)} \(g^{(2)}(0)\) as a function of the emitter number \(N\) and emitter-drive strength \(|\varepsilon_e|\). The blue dashed curve marks the optimal condition \(|\varepsilon_e|=\varepsilon_a^2/g_N\). The fixed parameters are \(\Delta_c=0\), \(g=6\gamma_c\), \(\varepsilon_a=0.1\gamma_c\), and \(\gamma_e=\gamma_c\).}
  \label{fig:optimal_parameters}
\end{figure}

We further investigate the control of PB through the drive amplitudes and phases. Figure~\ref{fig:optimal_parameters}(a) shows a two-dimensional map of \(g^{(2)}(0)\) as a function of \(\phi_e\) and \(\phi_a\). Regions with different antibunching strengths appear as alternating parallel bands, demonstrating that the blockade can be continuously tuned by varying only the drive phases. This behavior is fully consistent with the optimal phase-matching relation \(\phi_e=2(\phi_a+k\pi)\). Figure~\ref{fig:optimal_parameters}(b) shows \(g^{(2)}(0)\) as a function of the emitter number \(N\) and the collective emitter-drive strength \(|\varepsilon_e|\), for fixed cavity drive and individual coupling strength. The optimal emitter-drive amplitude decreases with increasing \(N\), implying that only a weak collective emitter drive is required to realize the optimal blockade in a many-emitter system. At the same time, the blockade is sensitive to amplitude matching: when the drive deviates from its optimal value, \(g^{(2)}(0)\) can increase substantially and can even become larger than in the purely collective scheme.

\subsection{Universal versus Unconventional Photon Blockade}
\label{sec:universal_vs_unconventional}

The universal PB scheme described above simultaneously exploits the single-photon resonance and two-path quantum interference. The two-photon Jaynes--Cummings model also supports another, purely interference-based mechanism, namely UPB. Its optimal conditions require a nonzero detuning \(\Delta_c=ng\) with \(n\neq0\), so that the system completely avoids the single-photon resonance and suppresses the two-photon state solely through destructive interference between different pathways. Previous comparisons in the single-emitter case showed that the two schemes yield nearly identical blockade strengths in the weak-coupling regime, with universal PB having only a modest advantage in the single-photon population \cite{37}. A natural question is whether collective enhancement acts equally on these two interference mechanisms in a multi-emitter system.

To answer this question, we first derive the optimal parameters for UPB in the collective two-photon Tavis--Cummings model. Unlike universal PB, UPB must operate away from the single-photon resonance, and we therefore set
\begin{equation}
\Delta_c=ng_N,
\qquad n\neq0,
\end{equation}
where \(n\) is a dimensionless real number. Within the Holstein--Primakoff approximation and the weak-excitation truncation, \(C_{20}\) is still given by Eq.~\eqref{eq:C20}. We impose \(C_{20}=0\), define the phase difference
\begin{equation}
\Delta_\phi=2\phi_a-\phi_e,
\end{equation}
and set the real and imaginary parts of the numerator in Eq.~\eqref{eq:C20} separately to zero. In the weak-coupling regime \(g_N\lesssim\gamma\), with \(\gamma_c=\gamma_e=\gamma\), this procedure gives
\begin{align}
\Delta_c={}&ng_N,\\
g_N={}&\frac{
\csc\Delta_\phi}{16n^2}
\Bigl[\sqrt{2\gamma^2n^2(-7+9\cos2\Delta_\phi)} \notag\\
&\hspace{4.5em}-2\gamma n\cos\Delta_\phi\Bigr],\\
\varepsilon_e={}&\frac{\varepsilon_a^2}{2\gamma^2}
\Bigl[
\sqrt{2\gamma^2n^2(-7+9\cos2\Delta_\phi)}\cot\Delta_\phi \notag\\
&\hspace{3.0em}
+\gamma n(5-3\cos2\Delta_\phi)\csc\Delta_\phi
\Bigr].
\end{align}
For parameters outside this weak-coupling regime, we obtain the optimal pair \((\varepsilon_e,\Delta_\phi)\) by numerically solving for the vanishing real and imaginary parts of \(C_{20}\). In this general case, \(g_N\) is treated as a prescribed system parameter and is no longer constrained by the analytical relation above; destructive interference can nevertheless be realized by appropriately adjusting the drive amplitude and phase. Importantly, the optimal detuning \(\Delta_c=ng_N\) of UPB undergoes a collective shift with emitter number and therefore moves progressively away from the cavity resonance as \(N\) increases.

\begin{figure}[t]
  \centering
  \includegraphics[width=1.0\columnwidth]{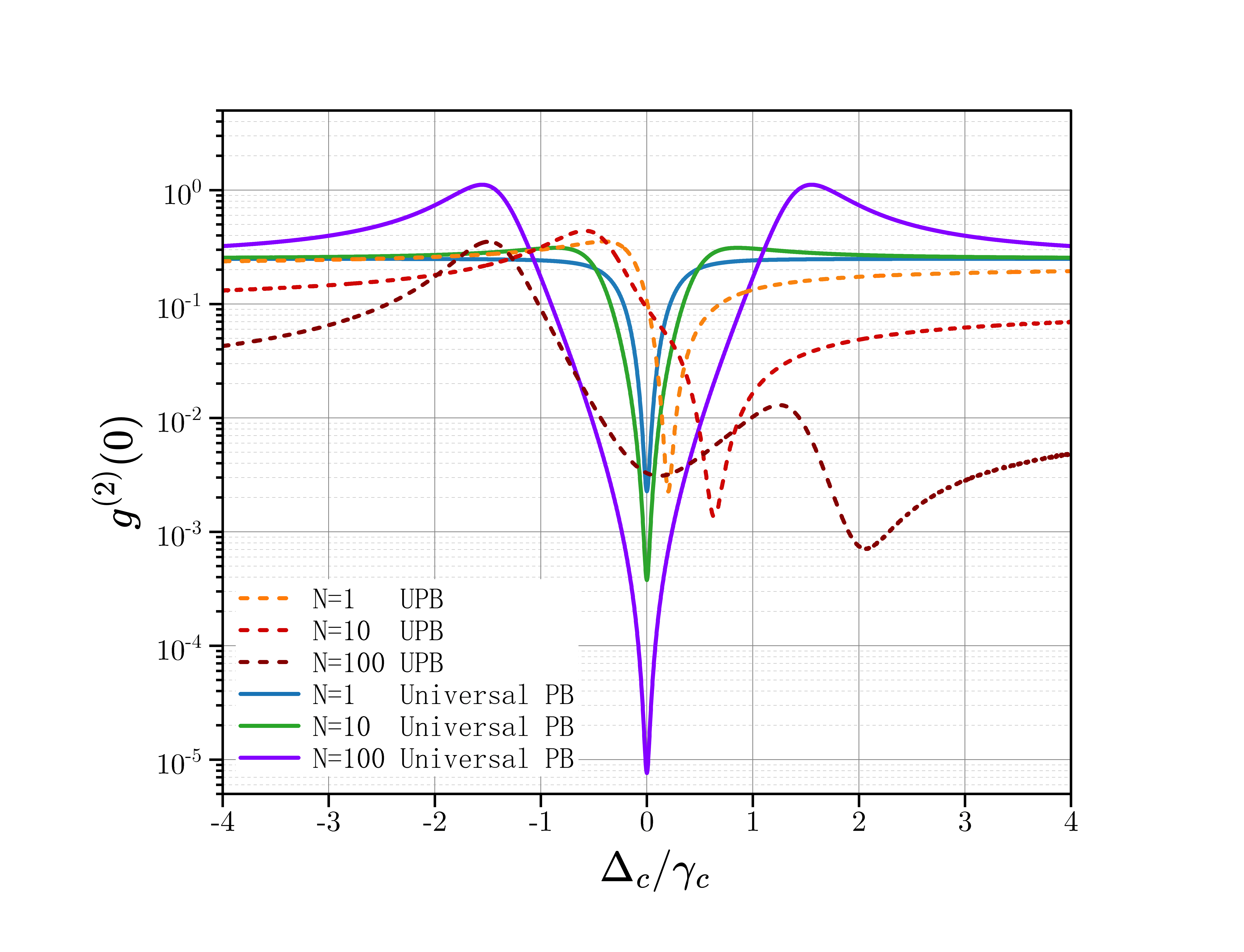}
  \caption{\textbf{Comparison between universal photon blockade (universal PB) and unconventional photon blockade (UPB).}
  Within the Holstein--Primakoff approximation, \(g^{(2)}(0)\) is plotted as a function of \(\Delta_c/\gamma_c\) for universal PB (solid curves) and UPB (dashed curves), with \(N=1,10,\) and \(100\). The fixed parameters are \(g/\gamma_c=0.2\), \(\varepsilon_a/\gamma_c=0.01\), \(\gamma_e=\gamma_c\), and \(\phi_a=0\). For universal PB, \(\varepsilon_e=-\varepsilon_a^2/g_N\) and \(\phi_e=0\), with the optimum occurring at \(\Delta_c=0\). For UPB, \(n=1\), and the drive amplitude and phase are determined numerically from the condition \(C_{20}=0\).}
  \label{fig:universal_vs_unconventional}
\end{figure}

Figure~\ref{fig:universal_vs_unconventional} shows \(g^{(2)}(0)\) as a function of the detuning \(\Delta_c\) for universal PB and UPB at different emitter numbers. For \(N=1\), their minima are nearly equal, consistent with the results for a single-emitter system. As \(N\) increases, however, the two schemes display markedly different trends. For universal PB, the minimum second-order correlation function approximately follows
\begin{equation}
g_{\min}^{(2)}(0)\propto\frac{1}{N^2},
\end{equation}
while the optimal detuning remains fixed at \(\Delta_c=0\). Although \(g^{(2)}(0)\) also decreases for UPB, the improvement is much smaller, and its optimal detuning moves increasingly far from resonance. The performance gap between the two schemes therefore widens with emitter number.

This difference can be understood in terms of both purity and brightness. In terms of purity, universal PB additionally benefits from the enhanced level anharmonicity associated with the collective coupling \(g_N\), so that \(g^{(2)}(0)\) decreases substantially as \(N\) increases. By contrast, UPB suppresses multiphoton excitation mainly through destructive interference between transition pathways, and an increase in collective coupling does not provide the same direct anharmonicity gain as in CPB. Its purity therefore improves more slowly. In terms of brightness, universal PB always operates at the resonance center and maintains a relatively high intracavity photon number. The optimal detuning of UPB moves away from resonance with increasing \(N\), which reduces the intracavity photon number at fixed drive strength. Thus, although the two schemes perform similarly for a single emitter, universal PB offers better scalability and greater promise for single-photon-source applications in multi-emitter systems.

\subsection{Control of Interference Blockade by Drive Strength and Spontaneous Emission}
\label{sec:drive_decay}

\begin{figure*}[t]
  \centering
  \includegraphics[width=1\linewidth]{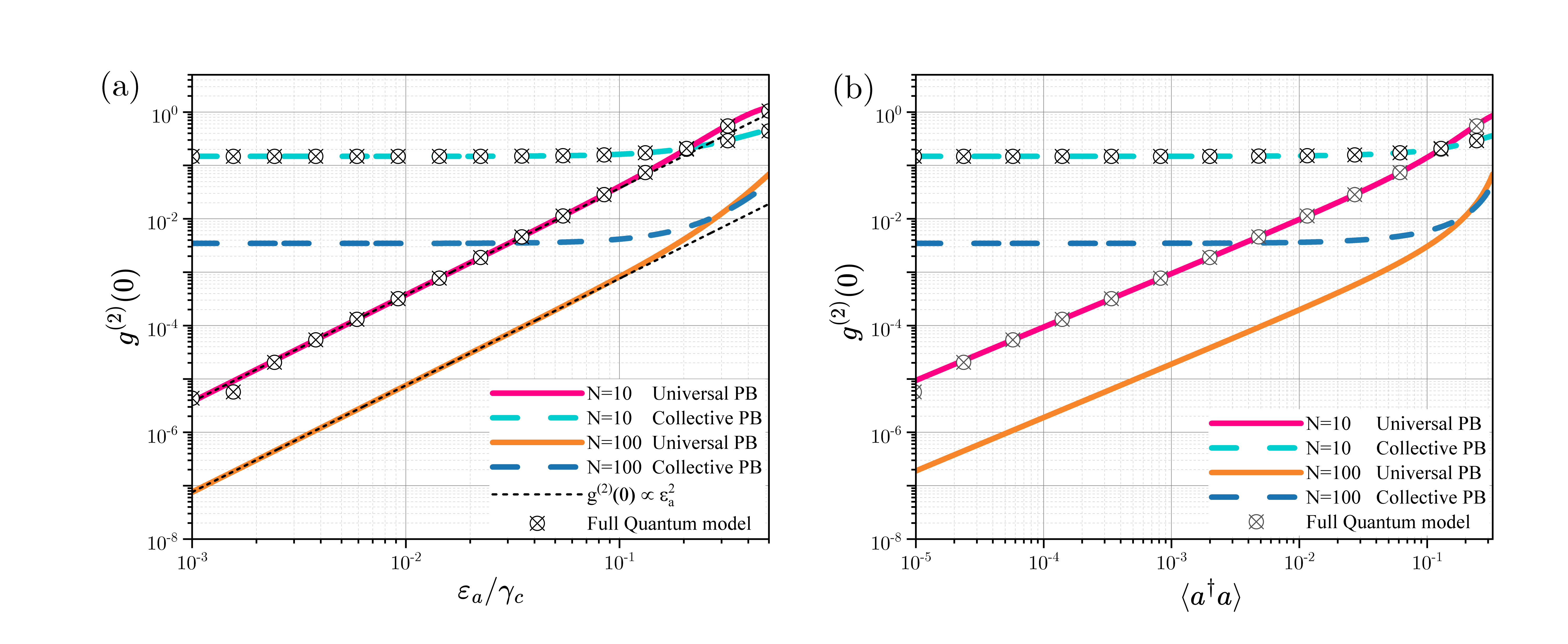}
  \caption{\textbf{Control of photon blockade by the drive strength.}
  \textbf{(a)} \(g^{(2)}(0)\) as a function of \(\varepsilon_a/\gamma_c\). The black dashed line indicates the reference scaling \(g^{(2)}(0)\propto\varepsilon_a^2\).
  \textbf{(b)} \(g^{(2)}(0)\) as a function of the mean intracavity photon number \(\langle a^\dagger a\rangle\), illustrating the purity--brightness trade-off. Solid and dashed curves correspond to \(N=100\) and \(N=10\), respectively, and the symbols are full-quantum simulation results. The fixed parameters are \(g/\gamma_c=0.2\) and \(\gamma_e=\gamma_c\), with the drives satisfying the optimal interference condition.}
  \label{fig:drive}
\end{figure*}

The drive strength plays a distinctive control role in universal PB. Figure~\ref{fig:drive}(a) shows \(g^{(2)}(0)\) as a function of the cavity-drive amplitude \(\varepsilon_a\) under the optimal conditions, together with the result for the purely collective scheme. For universal PB, the numerical results show that in the weak-drive regime
\begin{equation}
g^{(2)}(0)\propto\varepsilon_a^2.
\label{eq:drive_scaling}
\end{equation}
This quadratic scaling is independent of the emitter number, which changes only the proportionality coefficient of \(g^{(2)}(0)\). In contrast, \(g^{(2)}(0)\) for the purely collective scheme is nearly independent of \(\varepsilon_a\) throughout the weak-drive regime. This distinction reflects the different underlying mechanisms. The purely collective scheme relies on level anharmonicity, and its blockade performance is governed primarily by the effective coupling \(g_N\). Universal PB relies not only on level anharmonicity but also on the destructive-interference condition between distinct excitation pathways; higher-order excitation processes therefore cause its residual second-order correlation to depend on the drive strength.

Under the ideal optimal-interference condition, the \(N_{\mathrm{ex}}\leq2\) truncation gives \(C_{20}=0\) and thus predicts \(g^{(2)}(0)=0\). Both the Holstein--Primakoff simulations and the full-quantum master-equation results, however, exhibit the scaling in Eq.~\eqref{eq:drive_scaling}. Extending the wavefunction truncation to \(N_{\mathrm{ex}}\leq3\) shows that the residual contribution to the second-order correlation function arises predominantly from the three-photon state \(|3,0\rangle\). Although the two lowest-order pathways to the two-photon state cancel under the optimal conditions, a higher-order bypass pathway through \(|1,1\rangle\) leads to \(|3,0\rangle\) without passing through \(|2,0\rangle\) and is therefore not fully suppressed by the lowest-order interference condition. The corresponding probability-amplitude derivation is given in the Appendix.

At the same time, reducing the drive strength lowers the mean intracavity photon number \(\langle a^\dagger a\rangle\). Pursuing higher purity through the drive strength therefore sacrifices some brightness. Figure~\ref{fig:drive}(b) displays this purity--brightness trade-off. At the same intracavity photon number, \(g^{(2)}(0)\) is always lower for universal PB than for the purely collective scheme, and the purity difference becomes more pronounced as the brightness decreases. Thus, in addition to collective enhancement through increasing the emitter number, which introduces no extra brightness loss, the drive strength provides an independent and continuously tunable control parameter for the purity--brightness balance.

\begin{figure}[t]
  \centering
  \includegraphics[width=1\linewidth]{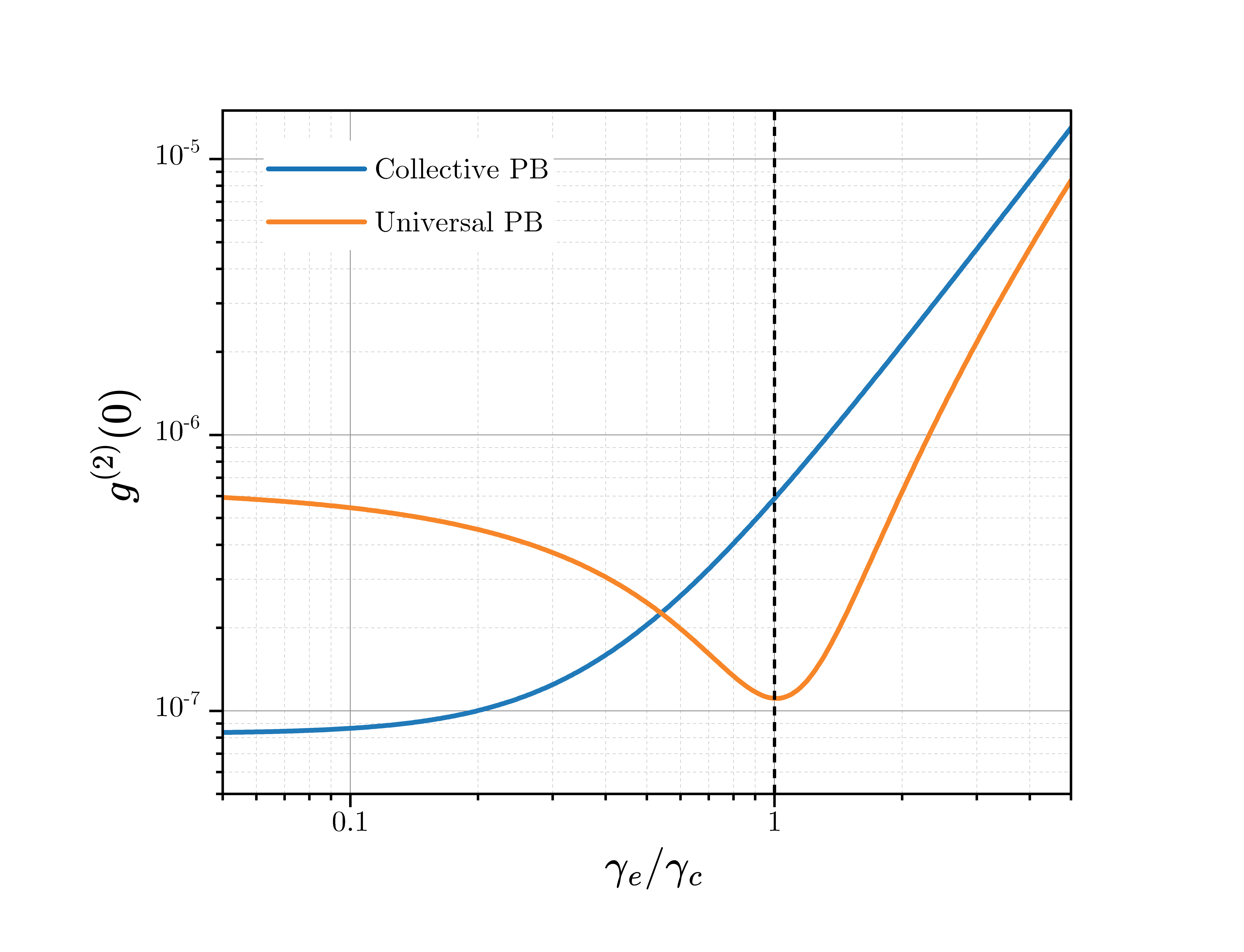}
  \caption{\textbf{Sensitivity to mismatched dissipation rates.}
  At the resonance point \(\Delta_c=0\), \(g^{(2)}(0)\) for universal PB and purely collective PB is plotted as a function of \(\gamma_e/\gamma_c\). The vertical dashed line marks the symmetric-dissipation condition \(\gamma_e=\gamma_c\).}
  \label{fig:decay}
\end{figure}

In realistic systems, the ratio of the emitter spontaneous-emission rate \(\gamma_e\) to the cavity decay rate \(\gamma_c\) depends on the experimental platform and need not satisfy the symmetric condition \(\gamma_e=\gamma_c\). It is therefore important to examine how deviations from this optimal assumption affect the scheme. Figure~\ref{fig:decay} shows \(g^{(2)}(0)\) at resonance as a function of \(\gamma_e/\gamma_c\) for the universal PB and purely collective schemes. At \(\gamma_e=\gamma_c\), universal PB is optimal and its \(g^{(2)}(0)\) is substantially lower than that of the purely collective scheme. Once \(\gamma_e\) departs from this symmetric condition, however, the performance of universal PB rapidly deteriorates, whereas \(g^{(2)}(0)\) for the purely collective scheme varies more gradually. The mismatch between the dissipation rates disrupts the balance between the two excitation pathways and prevents exact destructive interference. The purely collective scheme relies mainly on level anharmonicity and is therefore more robust against such a mismatch.

\section{Conclusion}

We have proposed and systematically investigated a collectively enhanced universal photon blockade (universal PB) scheme that extends destructive interference between two excitation pathways to a multi-emitter system and combines it with collective enhancement in the two-photon Tavis--Cummings model. By cross-validating full-quantum numerical simulations, the Holstein--Primakoff approximation, and a probability-amplitude analysis based on a non-Hermitian effective Hamiltonian, we established the theoretical framework of the scheme and elucidated its underlying physical mechanism. Our analytical treatment yielded the optimal conditions for universal PB, which combine the energy-resonance requirement of CPB with the phase- and amplitude-matching requirements for destructive interference between two pathways. The two-photon excitation is thereby strongly suppressed. Compared with the purely collective PB scheme, which relies only on level anharmonicity, the proposed scheme retains the benefit of collective enhancement while using pathway interference to further improve the single-photon purity. As the emitter number \(N\) increases, the antibunching improves continuously without requiring a reduction in drive strength or a shift away from the cavity resonance, and collective enhancement itself therefore introduces no additional brightness loss.More importantly, universal PB displays a more favorable collective scaling than UPB in a multi-emitter system. Its optimal operating point remains at the cavity resonance, and the minimum second-order correlation function follows \(g_{\min}^{(2)}(0)\propto N^{-2}\). By contrast, the optimal detuning of UPB moves away from resonance as \(N\) increases; its antibunching improves only modestly and is accompanied by a pronounced reduction in brightness. Thus, the two interference-based blockade mechanisms, which perform similarly for a single emitter, exhibit distinct collective scaling behaviors in a multi-emitter system.The proposed scheme also provides a form of drive tunability absent from the purely collective scheme. In the weak-drive regime, universal PB obeys \(g^{(2)}(0)\propto\varepsilon_a^2\), originating from higher-order bypass pathways through the three-excitation manifold that lead to the three-photon state after the lowest-order two-photon excitation pathways have canceled. Consequently, in addition to increasing the emitter number to obtain collective enhancement without an extra loss of brightness, one may further improve the single-photon purity by reducing the drive power, at the cost of some brightness. On the other hand, the scheme relies on precise pathway interference and is therefore more sensitive to mismatched dissipation rates. When \(\gamma_e\) deviates from \(\gamma_c\), the pathway balance is disrupted and \(g^{(2)}(0)\) rapidly increases, whereas the purely collective scheme based on level anharmonicity exhibits greater robustness.In summary, this work reveals the cooperative mechanism between many-body collective effects and destructive interference between two excitation pathways. It demonstrates the combined advantages of relaxing the individual-emitter strong-coupling requirement, exploiting collective enhancement to generate high-quality single-photon emission under relatively weak individual coupling, maintaining relatively high brightness, and providing drive tunability. It also clarifies the selective action of collective enhancement on different interference-based blockade mechanisms and the associated robustness cost. These results provide a theoretical basis for designing and optimizing scalable single-photon sources in multi-emitter cavity-QED systems.

\appendix

\section{Non-Hermitian Probability-Amplitude Solution under the \texorpdfstring{\(N_{\mathrm{ex}}\leq3\)}{Nex <= 3} Truncation}
\label{app:three_excitation}

The analytical derivation in the main text employed the \(N_{\mathrm{ex}}\leq2\) truncation and successfully yielded the optimal interference conditions for universal PB. Under this truncation, however, the optimal conditions give \(C_{20}=0\) and therefore predict \(g^{(2)}(0)=0\). This result differs from the full master-equation simulations, which give a finite \(g^{(2)}(0)\) that increases as \(\varepsilon_a^2\) in the weak-drive regime. Thus, although the \(N_{\mathrm{ex}}\leq2\) truncation is sufficient to derive the lowest-order optimal interference conditions, it cannot describe the residual second-order correlation after destructive interference. We therefore extend the wavefunction truncation to the subspace with total excitation number \(N_{\mathrm{ex}}\leq3\):
\begin{align}
|\psi\rangle={}&C_{00}|0,0\rangle+C_{10}|1,0\rangle
+C_{01}|0,1\rangle \notag\\
&+C_{20}|2,0\rangle+C_{11}|1,1\rangle+C_{30}|3,0\rangle.
\end{align}

Substituting this state into the non-Hermitian Schr\"odinger equation, imposing the steady-state conditions \(\dot C_{ij}=0\), and taking \(C_{00}\simeq1\), together with the resonance conditions \(\Delta_c=\Delta_e=0\), symmetric dissipation \(\gamma_c=\gamma_e=\gamma\), and the optimal interference conditions
\begin{equation}
\phi_a=\phi_e=0,
\qquad
\varepsilon_e=-\frac{\varepsilon_a^2}{g_N},
\end{equation}
gives the following system of five linear equations:
\begin{align}
AC_{10}+\sqrt2\varepsilon_aC_{20}+\varepsilon_eC_{11}&=-\varepsilon_a,\\
AC_{01}+\sqrt2g_NC_{20}+\varepsilon_aC_{11}&=-\varepsilon_e,\\
2AC_{20}+\sqrt2\varepsilon_aC_{10}+\sqrt2g_NC_{01}
+\sqrt3\varepsilon_aC_{30}&=0,\\
3AC_{30}+\sqrt3\varepsilon_aC_{20}+\sqrt6g_NC_{11}&=0,\\
2AC_{11}+\varepsilon_aC_{01}+\sqrt6g_NC_{30}
+\varepsilon_eC_{10}&=0,
\end{align}
where
\begin{equation}
A=-\frac{i\gamma}{2}.
\end{equation}
Defining
\begin{align}
L&=2A^2-3\varepsilon_a^2-2g_N^2,\\
K&=2A^2-2g_N^2-\varepsilon_a^2-\frac{\varepsilon_a^4}{g_N^2},\\
M&=\frac{\sqrt2\varepsilon_a(2g_N^2-\varepsilon_a^2)}{g_N},
\end{align}
the closed-form probability amplitudes under this three-excitation truncation and the weak-drive approximation \(C_{00}\simeq1\) are
\begin{align}
C_{11}&=-\frac{2\varepsilon_a^3}{g_N}\frac{L}{KL-M^2},\\
C_{20}&=\frac{M}{L}C_{11},\\
C_{30}&=-\frac{\sqrt3\varepsilon_aC_{20}+\sqrt6g_NC_{11}}{3A},\\
C_{10}&=\frac{-\varepsilon_a-\sqrt2\varepsilon_aC_{20}-\varepsilon_eC_{11}}{A},\\
C_{01}&=\frac{-\varepsilon_e-\sqrt2g_NC_{20}-\varepsilon_aC_{11}}{A}.
\end{align}

In the weak-drive limit \(\varepsilon_a\ll\gamma,g_N\), their leading orders are
\begin{align}
C_{10}&=\mathcal O(\varepsilon_a),&
C_{01}&=\mathcal O(\varepsilon_a^2),&
C_{11}&=\mathcal O(\varepsilon_a^3),\\
C_{30}&=\mathcal O(\varepsilon_a^3),&
C_{20}&=\mathcal O(\varepsilon_a^4).&&
\end{align}
Writing the wavefunction normalization as \(\mathcal N=\sum_{ij}|C_{ij}|^2\simeq1\), we obtain
\begin{align}
\langle a^\dagger a\rangle&=
\frac{|C_{10}|^2+2|C_{20}|^2+|C_{11}|^2+3|C_{30}|^2}{\mathcal N},\\
\langle a^{\dagger2}a^2\rangle&=
\frac{2|C_{20}|^2+6|C_{30}|^2}{\mathcal N}.
\end{align}
Because \(|C_{30}|\gg|C_{20}|\) in the weak-drive limit, the residual contribution to the second-order correlation function is dominated by the three-photon state \(|3,0\rangle\). Therefore,
\begin{equation}
g^{(2)}(0)\simeq
\frac{6|C_{30}|^2}{|C_{10}|^4}
\propto\frac{\varepsilon_a^6}{\varepsilon_a^4}
=\varepsilon_a^2.
\end{equation}

This result is consistent with the full master-equation simulations. The physical picture is as follows. Under the optimal interference conditions, the two lowest-order two-photon excitation pathways
\begin{equation}
|0,0\rangle\rightarrow|1,0\rangle\rightarrow|2,0\rangle,
\qquad
|0,0\rangle\rightarrow|0,1\rangle\rightarrow|2,0\rangle
\end{equation}
cancel exactly. The three-excitation manifold, however, introduces higher-order bypass pathways through \(|1,1\rangle\) to \(|3,0\rangle\), including
\begin{align}
|0,0\rangle&\xrightarrow{\varepsilon_e}|0,1\rangle
\xrightarrow{\varepsilon_a}|1,1\rangle
\xrightarrow{g_N}|3,0\rangle,\\
|0,0\rangle&\xrightarrow{\varepsilon_a}|1,0\rangle
\xrightarrow{\varepsilon_e}|1,1\rangle
\xrightarrow{g_N}|3,0\rangle.
\end{align}
Neither pathway passes through \(|2,0\rangle\), so they are not fully suppressed by the lowest-order interference condition. The resulting three-photon population gives the dominant residual contribution to the second-order correlation and ultimately determines the scaling \(g^{(2)}(0)\propto\varepsilon_a^2\).

\bibliography{ref}

@book{grabert2013single,
  title={Single charge tunneling: Coulomb blockade phenomena in nanostructures},
  author={Grabert, Hermann and Devoret, Michel H},
  volume={294},
  year={2013},
  publisher={Springer Science \& Business Media}
}

@article{9,
author = {Zubizarreta Casalengua, Eduardo and López Carreño, Juan Camilo and Laussy, Fabrice P. and Valle, Elena del},
title = {Conventional and unconventional photon statistics},
journal = {Laser Photonics Rev.},
year = {2020},
volume  = {14},
pages = {1900279},
doi = {10.1002/lpor.201900279},
url = {http://dx.doi.org/10.1002/lpor.201900279},
}

@article{10,
author = {Liu, Guan. Xin and Zhou, Wen. Jie and Gromyko, Dmitrii and Huang, Ding and Dong, Zhao. Gang and Liu, Ren. Ming and Zhu, Juan. Feng and Liu, Jing. Feng and Qiu, Cheng. Wei and Wu, Lin},
title = {Single-photon generation and manipulation in quantum nanophotonics},
journal = {Appl. Phys. Rev.},
year = {2025},
volume  = {12},
pages = {011308},
doi = {10.1063/5.0227350},
url = {http://dx.doi.org/10.1063/5.0227350},
}

@article{1,
author = {Gerace, Dario and Türeci, Hakan E. and Imamoglu, Atac and Giovannetti, Vittorio and Fazio, Rosario},
title = {The quantum-optical {J}osephson interferometer},
journal = {Nat. Phys.},
year = {2009},
volume  = {5},
pages = {281},
doi = {10.1038/nphys1223},
url = {http://dx.doi.org/10.1038/nphys1223},
}

@article{2,
author = {Rosenblum, Serge and Parkins, Scott and Dayan, Barak},
title = {Photon routing in cavity {QED}: Beyond the fundamental limit of photon blockade},
journal = {Phys. Rev. A},
year = {2011},
volume  = {84},
pages = {033854},
doi = {10.1103/physreva.84.033854},
url = {http://dx.doi.org/10.1103/physreva.84.033854},
}

@article{3,
author = {Shomroni, Itay and Rosenblum, Serge and Lovsky, Yulia and Bechler, Orel and Guendelman, Gabriel and Dayan, Barak},
title = {All-optical routing of single photons by a one-atom switch controlled by a single photon},
journal = {Science},
year = {2014},
volume  = {354},
pages = {903},
doi = {10.1126/science.1254699},
url = {http://dx.doi.org/10.1126/science.1254699},
}

@article{4,
author = {Chang, Darrick E. and Sørensen, Anders S. and Demler, Eugene A. and Lukin, Mikhail D.},
title = {A single-photon transistor using nanoscale surface plasmons},
journal = {Nat. Phys.},
year = {2007},
volume  = {3},
pages = {807},
doi = {10.1038/nphys708},
url = {http://dx.doi.org/10.1038/nphys708},
}

@article{5,
author = {Liao, Ren and Song, Ze. Rui and Ye, Gen. Sheng and Yu, Jian. Hao and Chang, Yue and Li, Lin},
title = {Nonlocal switch and transistor between single photons},
journal = {Phys. Rev. Lett.},
year = {2025},
volume  = {135},
pages = {260803},
doi = {10.1103/2sl5-7pbr},
url = {http://dx.doi.org/10.1103/2sl5-7pbr},
}

@article{6,
author = {Sayrin, Clément and Junge, Christian and Mitsch, Rudolf and Albrecht, Bernhard and O’Shea, Danny and Schneeweiss, Philipp and Volz, Jürgen and Rauschenbeutel, Arno},
title = {Nanophotonic optical isolator controlled by the internal state of cold atoms},
journal = {Phys. Rev. X},
year = {2015},
volume  = {5},
pages = {041036},
doi = {10.1103/physrevx.5.041036},
url = {http://dx.doi.org/10.1103/physrevx.5.041036},
}

@article{7,
author = {Tang, Lei and Tang, Jiang. Shan and Zhang, Wei. Dong and Lu, Guo. Wei and Zhang, Han and Zhang, Yong and Xia, Ke. Yu and Xiao, Min},
title = {On-chip chiral single-photon interface: Isolation and unidirectional emission},
journal = {Phys. Rev. A},
year = {2019},
volume  = {99},
pages = {043833},
doi = {10.1103/physreva.99.043833},
url = {http://dx.doi.org/10.1103/physreva.99.043833},
}

@article{11,
author = {Imamoğlu, A. and Schmidt, H. and Woods, G. and Deutsch, M.},
title = {Strongly interacting photons in a nonlinear cavity},
journal = {Phys. Rev. Lett.},
year = {1997},
volume  = {79},
pages = {1467},
doi = {10.1103/PhysRevLett.79.1467},
url = {https://link.aps.org/doi/10.1103/PhysRevLett.79.1467},
}

@article{12,
author = {Birnbaum, K. M. and Boca, A. and Miller, R. and Boozer, A. D. and Northup, T. E. and Kimble, H. J.},
title = {Photon blockade in an optical cavity with one trapped atom},
journal = {Nature},
year = {2005},
volume  = {436},
pages = {87},
doi = {10.1038/nature03804},
url = {http://dx.doi.org/10.1038/nature03804},
}

@article{13,
author = {Faraon, Andrei and Fushman, Ilya and Englund, Dirk and Stoltz, Nick and Petroff, Pierre and Vučković, Jelena},
title = {Coherent generation of non-classical light on a chip via photon-induced tunnelling and blockade},
journal = {Nat. Phys.},
year = {2008},
volume  = {4},
pages = {859},
doi = {10.1038/nphys1078},
url = {http://dx.doi.org/10.1038/nphys1078},
}

@article{15,
author = {Li, Zi. Geng and Li, Xiao. Miao and Zhong, Xiao. Lan},
title = {Strong photon blockade in an all-fiber emitter-cavity quantum electrodynamics system},
journal = {Phys. Rev. A},
year = {2021},
volume  = {103},
pages = {043724},
doi = {10.1103/physreva.103.043724},
url = {http://dx.doi.org/10.1103/physreva.103.043724},
}

@article{16,
author = {Tang, Jing and Deng, Yuan. Gang},
title = {Tunable multiphoton bundles emission in a {K}err-type two-photon {J}aynes-{C}ummings model},
journal = {Phys. Rev. Res.},
year = {2024},
volume  = {6},
pages = {033247},
doi = {10.1103/physrevresearch.6.033247},
url = {http://dx.doi.org/10.1103/physrevresearch.6.033247},
}

@article{17,
author = {Bin, Qian and Wu, Ying and Gao, Jin. Hua and Chen, Ai. Xi and Nori, Franco and Lü, Xin. You},
title = {Cavity {QED} based on strongly localized modes: Exponentially enhancing single-atom cooperativity},
journal = {Phys. Rev. Lett.},
year = {2025},
volume  = {135},
pages = {103602},
doi = {10.1103/7frd-pf1m},
url = {http://dx.doi.org/10.1103/7frd-pf1m},
}

@article{18,
author = {Liew, T. C. H. and Savona, V.},
title = {Single photons from coupled quantum modes.},
journal = {Phys. Rev. Lett.},
volume  = {104},
year = {2010},
pages = {183601},
doi = {10.1103/physrevlett.104.183601},
url = {http://dx.doi.org/10.1103/physrevlett.104.183601},
}

@article{19,
author = {Bamba, Motoaki and Imamoğlu, Atac and Carusotto, Iacopo and Ciuti, Cristiano},
title = {Origin of strong photon antibunching in weakly nonlinear photonic molecules},
journal = {Phys. Rev. A},
volume  = {83},
year = {2011},
pages = {021802},
doi = {10.1103/physreva.83.021802},
url = {http://dx.doi.org/10.1103/physreva.83.021802},
}

@article{21,
author = {Zhang, Wei and Hou, Rui and Wang, Tie and Liu, Shu. Tian and Zhang, Shou and Wang, Hong. Fu},
title = {Simultaneous nonreciprocal photon blockade via directional parametric amplification},
journal = {Phys. Rev. A},
year = {2024},
volume  = {110},
pages = {023723},
doi = {10.1103/physreva.110.023723},
url = {http://dx.doi.org/10.1103/physreva.110.023723},
}

@article{27,
author = {Snijders, H. J. and Frey, J. A. and Norman, J. and Flayac, H. and Savona, V. and Gossard, A. C. and Bowers, J. E. and van Exter, M. P. and Bouwmeester, D. and Löffler, W.},
title = {Observation of the unconventional photon blockade},
journal = {Phys. Rev. Lett.},
year = {2018},
volume  = {121},
pages = {043601},
doi = {10.1103/physrevlett.121.043601},
url = {http://dx.doi.org/10.1103/physrevlett.121.043601},
}

@article{31,
author = {Vaneph, Cyril and Morvan, Alexis and Aiello, Gianluca and Féchant, Mathieu and Aprili, Marco and Gabelli, Julien and Estève, Jérôme},
title = {Observation of the unconventional photon blockade in the microwave domain},
journal = {Phys. Rev. Lett.},
volume  = {121},
year = {2018},
pages = {043602},
doi = {10.1103/physrevlett.121.043602},
url = {http://dx.doi.org/10.1103/physrevlett.121.043602},
}

@article{37,
author = {Zhou, Yan. Hui and Liu, Tong and Su, Qi. Ping and Zhang, Xing. Yuan and Wu, Qi. Cheng and Chen, Dong. Xu and Shi, Zhi. Cheng and Shen, H. Z. and Yang, Chui. Ping},
title = {Universal photon blockade},
journal = {Phys. Rev. Lett.},
year = {2025},
volume = {134},
pages = {183601},
doi = {10.1103/physrevlett.134.183601},
url = {http://dx.doi.org/10.1103/physrevlett.134.183601},
}

@article{38,
author = {Zhang, Wei and Hou, Rui and Liu, Shu. Tian and Zhang, Shou and Wang, Hong. Fu},
title = {Universal nonreciprocal photon blockade},
journal = {Sci. China Phys. Mech. Astron.},
year = {2026},
volume = {69},
pages = {240313},
doi = {10.1007/s11433-025-2867-5},
url = {http://dx.doi.org/10.1007/s11433-025-2867-5},
}

@article{39,
author = {Lv, Guo. Hua and Hou, Rui and Zhou, Yan. Hui and Han, Xue and Wang, Hong. Fu and Zhang, Shou},
title = {Universal photon blockades via parametric amplification in second-order nonlinearly coupled cavities},
journal = {Opt. Express},
year = {2026},
volume = {34},
pages = {412},
doi = {10.1364/oe.582699},
url = {http://dx.doi.org/10.1364/oe.582699},
}

@article{372,
author = {Zhou, Yan. Hui and Liu, Tong and Su, Qi. Ping and Wu, Qi. Cheng and Guo, Bao. Qing and Shi, Zhi. Cheng and Shen, H. Z. and Yang, Chui. Ping},
title = {Universal photon blockade in two coupled cavities with {K}err nonlinearities},
journal = {Phys. Rev. A},
year = {2026},
volume = {113},
pages = {033722},
doi = {10.1103/vdjq-wyv8},
url = {https://link.aps.org/doi/10.1103/vdjq-wyv8},
}

@misc{chang2026,
author = {Guohao Chang and Hunduz Halemjan and Shangyun Liu and Raziya Anwar and Ahmad Abliz},
title = {Robust universal photon blockade in a bimodal {J}aynes-{C}ummings model via {K}err nonlinearity},
year = {2026},
eprint={2604.03838},
archivePrefix={arXiv},
primaryClass={quant-ph},
url = {https://arxiv.org/abs/2604.03838},
}

@misc{dong2026,
author = {Hai-Tao Dong and Meng-Long Song and Si-Yu Zhang and Xue-Ke Song and Liu Ye and Dong Wang},
title = {Universal photon blockade via two-photon light-matter interaction at chiral exceptional points},
year = {2026},
eprint={2606.18756},
archivePrefix={arXiv},
primaryClass={quant-ph},
url = {https://arxiv.org/abs/2606.18756},
}

@article{universal2026,
author = {Sun, Cheng-Ze and Yan, Wei-Bin and Yin, Rui-Bin and Man, Zhong-Xiao and Xia, Yun-Jie and Zhang, Ying-Jie and Cai, Qing-Yu},
title = {Universal nonreciprocal photon blockade},
journal = {Phys. Rev. A},
year = {2026},
volume = {113},
pages = {033720},
doi = {10.1103/cv14-4pdc},
url = {http://dx.doi.org/10.1103/cv14-4pdc},
}

@article{46,
author = {Johansson, J.R. and Nation, P.D. and Nori, Franco},
title = {{QuTiP}: An open-source {P}ython framework for the dynamics of open quantum systems},
journal = {Comput. Phys. Commun.},
year = {2012},
volume = {183},
pages = {1760},
doi = {10.1016/j.cpc.2012.02.021},
url = {http://dx.doi.org/10.1016/j.cpc.2012.02.021},
}

@article{47,
author = {Johansson, J.R. and Nation, P.D. and Nori, Franco},
title = {{QuTiP} 2: A {P}ython framework for the dynamics of open quantum systems},
journal = {Comput. Phys. Commun.},
year = {2013},
volume = {184},
pages = {1234},
doi = {10.1016/j.cpc.2012.11.019},
url = {http://dx.doi.org/10.1016/j.cpc.2012.11.019},
}

@article{qutip,
author = {Shammah, Nathan and Ahmed, Shahnawaz and Lambert, Neill and De Liberato, Simone and Nori, Franco},
title = {Open quantum systems with local and collective incoherent processes: Efficient numerical simulations using permutational invariance},
journal = {Phys. Rev. A},
year = {2018},
volume = {98},
pages = {063815},
doi = {10.1103/physreva.98.063815},
url = {http://dx.doi.org/10.1103/physreva.98.063815},
}

@article{45,
author = {Liu, Yu. Long and Wang, Guan. Zhong and Liu, Yu. Xi and Nori, Franco},
title = {Mode coupling and photon antibunching in a bimodal cavity containing a dipole quantum emitter},
journal = {Phys. Rev. A},
year = {2016},
volume = {93},
pages = {013856},
doi = {10.1103/physreva.93.013856},
url = {http://dx.doi.org/10.1103/physreva.93.013856},
}

@article{48,
author = {Myilswamy, Karthik V. and Seshadri, Suparna and Lu, Hsuan. Hao and Alshaykh, Mohammed S. and Liu, Jun. Qiu and Kippenberg, Tobias J. and Weiner, Andrew M. and Lukens, Joseph M.},
title = {Time-resolved {Hanbury Brown–Twiss} interferometry of on-chip biphoton frequency combs using {V}ernier phase modulation},
journal = {Phys. Rev. Appl.},
year = {2023},
volume = {19},
pages = {034019},
doi = {10.1103/physrevapplied.19.034019},
url = {http://dx.doi.org/10.1103/physrevapplied.19.034019},
}

@article{HBT1956,
author = {Hanbury Brown, R. and Twiss, R. Q.},
title = {Correlation between photons in two coherent beams of light},
journal = {Nature},
year = {1956},
volume = {177},
pages = {27},
doi = {10.1038/177027a0},
url = {https://doi.org/10.1038/177027a0},
}

@article{realized1,
author = {Villas-Boas, Celso J. and Rossatto, Daniel Z.},
title = {Multiphoton {J}aynes-{C}ummings model: Arbitrary rotations in {F}ock space and quantum filters},
journal = {Phys. Rev. Lett.},
year = {2019},
volume = {122},
pages = {123604},
doi = {10.1103/physrevlett.122.123604},
url = {http://dx.doi.org/10.1103/physrevlett.122.123604},
}

@article{realized2,
author = {Felicetti, S. and Rossatto, D. Z. and Rico, E. and Solano, E. and Forn-Díaz, P.},
title = {Two-photon quantum {R}abi model with superconducting circuits},
journal = {Phys. Rev. A},
year = {2018},
volume = {97},
pages = {013851},
doi = {10.1103/physreva.97.013851},
url = {http://dx.doi.org/10.1103/physreva.97.013851},
}

@article{realized3,
author = {Zou, Fen and Zhang, Xiao-Ya and Xu, Xun-Wei and Huang, Jin-Feng and Liao, Jie-Qiao},
title = {Multiphoton blockade in the two-photon Jaynes-Cummings model},
journal = {Phys. Rev. A},
year = {2020},
volume = {102},
pages = {053710},
doi = {10.1103/physreva.102.053710},
url = {http://dx.doi.org/10.1103/physreva.102.053710},
}

@article{realized4,
author = {Li, Hai-Ji and Fan, Li-Bao and Ma, Shan and Liao, Jie-Qiao and Shu, Chuan-Cun},
title = {Exploring photon blockade in a two-photon Jaynes-Cummings model with atom and cavity drivings},
journal = {Phys. Rev. A},
year = {2024},
volume = {110},
pages = {043707},
doi = {10.1103/physreva.110.043707},
url = {http://dx.doi.org/10.1103/physreva.110.043707},
}

@article{realized5,
  title = {Nonlinear {J}aynes-{C}ummings dynamics of a trapped ion},
  author = {Vogel, W. and Filho, R. L. de Matos},
  journal = {Phys. Rev. A},
  volume = {52},
  pages = {4214},
  year = {1995},
  doi = {10.1103/PhysRevA.52.4214},
  url = {https://link.aps.org/doi/10.1103/PhysRevA.52.4214}
}

@article{realized6,
  title = {Generation of nonclassical motional states of a trapped atom},
  author = {Meekhof, D. M. and Monroe, C. and King, B. E. and Itano, W. M. and Wineland, D. J.},
  journal = {Phys. Rev. Lett.},
  volume = {76},
  pages = {1796},
  year = {1996},
  doi = {10.1103/PhysRevLett.76.1796},
  url = {https://link.aps.org/doi/10.1103/PhysRevLett.76.1796}
}

@article{realized7,
  title = {Realization of a continuous-wave, two-photon optical laser},
  author = {Gauthier, Daniel J. and Wu, Qilin and Morin, S. E. and Mossberg, T. W.},
  journal = {Phys. Rev. Lett.},
  volume = {68},
  pages = {464},
  year = {1992},
  doi = {10.1103/PhysRevLett.68.464},
  url = {https://link.aps.org/doi/10.1103/PhysRevLett.68.464}
}

@article{realized8,
  title = {Two-photon lasing by a superconducting qubit},
  author = {Neilinger, P. and Reh\'ak, M. and Grajcar, M. and Oelsner, G. and H\"ubner, U. and Il'ichev, E.},
  journal = {Phys. Rev. B},
  volume = {91},
  pages = {104516},
  year = {2015},
  doi = {10.1103/PhysRevB.91.104516},
  url = {https://link.aps.org/doi/10.1103/PhysRevB.91.104516}
}

@article{realized9,
  title = {Multiphoton quantum {R}abi oscillations in ultrastrong cavity {QED}},
  author = {Garziano, Luigi and Stassi, Roberto and Macr\`{\i}, Vincenzo and Kockum, Anton Frisk and Savasta, Salvatore and Nori, Franco},
  journal = {Phys. Rev. A},
  volume = {92},
  pages = {063830},
   year = {2015},
  doi = {10.1103/PhysRevA.92.063830},
  url = {https://link.aps.org/doi/10.1103/PhysRevA.92.063830}
}

@article{realized10,
  title = {Ultrastrong-coupling regime of nondipolar light-matter interactions},
  author = {Felicetti, Simone and Hwang, Myung-Joong and Le Boit\'e, Alexandre},
  journal = {Phys. Rev. A},
  volume = {98},
  pages = {053859},
  year = {2018},
  doi = {10.1103/PhysRevA.98.053859},
  url = {https://link.aps.org/doi/10.1103/PhysRevA.98.053859}
}

@article{Kirton2017PRL,
  author = {Peter Kirton and Jonathan Keeling},
  title = {Suppressing and restoring the {D}icke superradiance transition by dephasing and decay},
  journal = {Phys. Rev. Lett.},
  volume = {118},
  number = {12},
  pages = {123602},
  year = {2017},
  publisher = {American Physical Society},
  doi = {10.1103/PhysRevLett.118.123602}
}

@article{Kirton2018Superradiant,
  title={Superradiant and lasing states in driven-dissipative Dicke models},
  author={Kirton, Peter and Keeling, Jonathan},
  journal={New J. Phys.},
  volume={20},
  number={1},
  pages={015009},
  year={2018},
  publisher={IOP Publishing},
  doi={https://iopscience.iop.org/article/10.1088/1367-2630/aaa11d}
}

@article{phase,
author = {Dai, Kunzhe and Wu, Haiteng and Zhao, Peng and Li, Mengmeng and Liu, Qiang and Xue, Guangming and Tan, Xinsheng and Yu, Haifeng and Yu, Yang},
title = {Quantum simulation of the general semi-classical {R}abi model in regimes of arbitrarily strong driving},
journal = {Appl. Phys. Lett.},
year = {2017},
volume = {111},
pages = {242601},
doi = {10.1063/1.5006745},
url = {http://dx.doi.org/10.1063/1.5006745},
}

@article{Collective,
  title = {Collective enhancement of photon blockade via two-photon interactions},
  author = {Dong, Lijuan and Shah, Aanal Jayesh and Kirton, Peter and Alaeian, Hadiseh and Felicetti, Simone},
  journal = {PRX Quantum},
  volume = {7},
  pages = {020349},
  year = {2026},
  doi = {10.1103/l477-z8jr},
  url = {https://link.aps.org/doi/10.1103/l477-z8jr}
}

@article{down1,
  title = {Biexciton-mediated superradiant photon blockade},
  author = {Poshakinskiy, Alexander V. and Poddubny, Alexander N.},
  journal = {Phys. Rev. A},
  volume = {93},
  pages = {033856},
  year = {2016},
  doi = {10.1103/PhysRevA.93.033856},
  url = {https://link.aps.org/doi/10.1103/PhysRevA.93.033856}
}

@article{down2,
  title = {Thermal, Quantum Antibunching and Lasing Thresholds from Single Emitters to Macroscopic Devices},
  author = {Carroll, Mark Anthony and D'Alessandro, Giampaolo and Lippi, Gian Luca and Oppo, Gian-Luca and Papoff, Francesco},
  journal = {Phys. Rev. Lett.},
  volume = {126},
  pages = {063902},
  year = {2021},
  doi = {10.1103/PhysRevLett.126.063902},
  url = {https://link.aps.org/doi/10.1103/PhysRevLett.126.063902}
}

@article{down3,
  title = {Photon blockade and single-photon generation with multiple quantum emitters},
  author = {Chen, Mingyuan and Tang, Jiangshan and Tang, Lei and Wu, Haodong and Xia, Keyu},
  journal = {Phys. Rev. Res.},
  volume = {4},
  pages = {033083},
  year = {2022},
  doi = {10.1103/PhysRevResearch.4.033083},
  url = {https://link.aps.org/doi/10.1103/PhysRevResearch.4.033083}
}

@article{2019,
  title = {Photon Blockade in Weakly Driven Cavity Quantum Electrodynamics Systems with Many Emitters},
  author = {Trivedi, Rahul and Radulaski, Marina and Fischer, Kevin A. and Fan, Shanhui and Vu\ifmmode \check{c}\else \v{c}\fi{}kovi\ifmmode \acute{c}\else \'{c}\fi{}, Jelena},
  journal = {Phys. Rev. Lett.},
  volume = {122},
  pages = {243602},
  year = {2019},
  doi = {10.1103/PhysRevLett.122.243602},
  url = {https://link.aps.org/doi/10.1103/PhysRevLett.122.243602}
}
\end{document}